\documentclass{article}
\usepackage{INTERSPEECH2019}

\title{Salient Speech Representations Based on Cloned Networks}
\name{W. Bastiaan Kleijn,$^{1,2}$
  Felicia S. C.  Lim,$^{1}$
      Michael Chinen,$^{1}$  
      Jan Skoglund$^{1}$
}
\address{$^{1}$Google LLC., San Francisco, CA\\
$^{2}$School of Engineering and Computer Science, Victoria University of 
  Wellington, New Zealand 
}
\email{bastiaan.kleijn@ecs.vuw.ac.nz, flim@google.com, mchinen@google.com, jks@google.com}
\usepackage{amsfonts,amsmath}
\usepackage{xcolor}
\usepackage{subcaption}
\usepackage{graphicx}
\usepackage{cite}
\usepackage[title]{appendix}
\usepackage[ruled]{algorithm2e}
\usepackage{slashbox}
\usepackage{verbatim}
\usepackage{enumitem}

\graphicspath{{figures/}}

\begin{document}
\ninept
\maketitle
\begin{abstract}
We define \textit{salient} features as features that are shared by signals that are
defined as being \textit{equivalent} by a system designer. The definition allows the
designer to contribute qualitative information. We aim to find salient features that are
useful as conditioning for generative networks. We extract salient features by jointly
training a set of clones of an encoder network. Each network clone
receives as input a different signal from a set of equivalent signals. The objective
function encourages the network clones to map 
their input into a set of features that is identical across the clones. It additionally
encourages feature independence and, optionally, reconstruction of a desired target signal by a
decoder.  As an application, we train a  
system that extracts a time-sequence of feature vectors of speech and uses it as a
conditioning of a WaveNet generative system, facilitating both coding and
enhancement. 
\end{abstract}

\noindent\textbf{Index Terms}: Features, representation, speech, Siamese network, generative network, maximum mean discrepancy.
\section{Introduction} 
\label{s:introduction}

In speech processing, as well as other applications, it is valuable to find a meaningful
representation that summarizes 
the salient attributes of a signal.  We define a salient feature set as a feature set that
is shared by signals that are judged to be equivalent by a user. This definition allows
the user to contribute \textit{qualitative} knowledge to a feature extraction
procedure, a weak form of supervision. We aim to use this paradigm to
extract feature sets that can be used as 
conditioning for generative networks. A successful salient feature set can be used for
the storage or transmission of signals, or for the manipulation of signal
attributes. Examples are the robust coding of speech, changing the identity of a talker,
and resynthesizing a speech signal without noise.

Autoencoders provide perhaps the most natural approach towards finding a meaningful
representation, although other approaches such as generative adversarial networks can also be used,
e.g., \cite{chen2016infogan}. Autoencoders map an input signal to a set of latent variables, the 
\textit{representation}, with an \textit{encoder} and map the latent variables to the output with a
\textit{decoder}. In recent years, research has focused on various forms of the  
\textit{variational} autoencoder  (VAE) \cite{kingma2013auto,rezende2014stochastic}. The
primary goal of the original VAE was to create a generative system (the decoder) and the
encoder arose from the need for a practical training algorithm.  Because the encoder has a
stochastic nature, the latent representation is expected to have a smooth relation with the
input.
%
The original VAE formulation was found to ignore the input in many cases, making the latent
representation not meaningful. Numerous methods have been proposed to address this
problem, referred to as \textit{information preference property}  or \textit{posterior
collapse}, e.g.,
\cite{chen2016variational,zhao2017infovae,alemi2017information,PhuongMutualA,Huszar2017Ismax,braithwaite2018bounded}.    

In addition to signal features varying smoothly over the latent support space, it is
desirable that these features are disentangled. In disentangled representations, different
latent dimensions exclusively control different factors of variation of the data (e.g., location,
orientation) \cite{bengio2013representation,louizos2015variational}.
$\beta$-VAE \cite{higgins2017beta} re-weights the components of the VAE
objective function, which encourages disentanglement at the cost of a lower reconstruction accuracy
\cite{burgess2018understanding, esmaeili2018structured,  kim2018disentangling}. 
$\beta$-VAE can be interpreted \cite{burgess2018understanding} as approximation to the
information bottleneck (IB)
\cite{tishby2000information} and hence is related to \cite{alemi2016deep}. Other approaches
explicitly encourage independence by means of maximum mean discrepancy \cite{ louizos2015variational}
and, recently, by means of total correlation (the Kulback-Leibler divergence between the joint
distribution and the product of the marginal distributions)
\cite{kim2018disentangling}. 

Conventional VAEs are not aware of saliency as we defined it.  VAEs
commonly use a maximum-likelihood objective combined with a Gaussian assumption for
reconstruction accuracy, which is equivalent to a straight squared error criterion. The
network is not informed about the perceptual importance of features to a user. Hence, a VAE-based
representation may include features that are perceptually irrelevant but require a
significant rate allocation. Indeed,\cite{kleijn2017wavenet} shows that most information about
the signal waveform is perceptually not relevant.

Salient information has been separated in VAEs using semi-supervised learning that separates out 
nuisance variables and class information from the latent representation
\cite{louizos2015variational}. This approach requires explicit knowledge of these
variables for at least part of the database.

The IB method suggests a method for extracting features that are salient based on our definition. The IB finds a
representation that is maximally informative about a target output, given a bound on the
information available about the input.  The approach was illustrated with a
sequence of phonemes as target output \cite{hecht2005extraction}. By training a network to
extract features from a signal that are maximally informative about an equivalent signal,
salient features can be extracted. However, only pairs of equivalent signals can be used for each training instance. 

Our contribution is a method to extract from a signal a sequence of salient feature vectors. It is based on the Siamese network structure
\cite{bromley1994signature,hoffer2015deep,kamper2016deep}, but can be endowed with a
decoder. During training, clones of an encoder network with
identical weights are encouraged to map equivalent signals to shared feature vectors. Our
objective function encourages i) that the feature vectors are  
identical across the clones, ii) that the features are independent, iii) optionally, that
the feature set can be mapped to a
shared target signal.

We applied the new method to the extraction of robust features for speech coding and
enhancement. As equivalent signals we used various distorted speech signals and for the
optional  decoder we use a clean signal as target. Including a target signal enhances
the clarity of the linguistic meaning of an already natural-sounding signal. 
Experiments with WaveNet \cite{WaveNet:2016a}  confirmed that the resulting feature set
performs significantly better in terms of robustness to noise than a conventional feature set.


\section{Extracting Salient Features}
\label{s:siamese}
In this section we first motivate and describe the basic extraction network in
\ref{s:encoder}. We then describe clone-based training procedures for the method in 
\ref{s:clones}.  We denote random variables (RVs) with upper case and realizations with
  lower case font.

\subsection{The encoder network}
\label{s:encoder} 
Our objective is to extract salient representations that can be used as conditioning for
generative models of speech. As noted in section \ref{s:introduction}, we define salient features as features
that are shared between signals that are deemed \textit{equivalent} by users. From 
each signal within a set of equivalent signals, the encoder should extract
feature sequences with numerically similar values.

The encoder is a map from an input vector $x$ to a salient
feature vector $z$. A sequence is obtained by repeating this process.  For a meaningful
feature vector $z$, we desire the map to have three basic attributes:
\begin{enumerate}[nosep]
\item Signals that are equivalent result in (almost) identical features.  
\item The map is smooth. Different regions of feature  space can be identified with
  different signal attributes, rendering the representation 
  \textit{meaningful}. 
\item\label{att:2} The components have a prescribed variance and are independent. Ideally they are distentangled: the
  system then discovers a  'natural' set of independent features corresponding to
    ground-truth factors \cite{bengio2013representation}. 
\end{enumerate}

The first desired attribute requires a surjective mapping from the
speech signal to a salient feature set. We show it can be obtained by training clones of
a network with identical weights to output the same features for equivalent signals. 

The second desired attribute can be obtained  with
various strategies. A first approach is to simply restrict the mapping to be smooth
deterministic (e.g., Lipschitz smooth). A second approach is to use a stochastic map
during training. To see this, consider a feature variable of the form $Z= f(X) +
\epsilon$, where $f(\cdot)$ is a deterministic map and $\epsilon\sim 
 p_\epsilon$. Then $Z$ is more informative about the input $X$ if the map $f(\cdot)$ is smooth.
Such a probabilistic-mapping approach to obtaining a
meaningful representation is employed by VAEs, e.g.,
\cite{kingma2013auto,rezende2014stochastic,braithwaite2018bounded} and this proven 
strategy will be employed here, also because it aids with the third attribute.  While
the probabilistic mapping is used to encourage smoothness of the mapping during training,
it can be removed in the final deployment of the  resulting system for enhanced
fidelity. Then, the mapping from speech to features is deterministic at inference.

The third desired attribute of the map facilitates interpretation, coding, and
manipulation. Distentanglement of the ground-truth factors is challenging, as a
mapping from speech to salient features can map a continuously 
distributed random speecph vector into a random feature vector with any desired
distribution. Without loss of generality, let us consider unit-variance
factors. Separation of these factors becomes tractable in the context of equivalent
signals, since their variance across an equivalent signal set is, in general,
distinct from each other. Hence a 'natural' set of features can be discovered.


We implement the probabilistic encoder as follows. Let $\phi$ describe the parameter set
of the encoder network. We then construct a network of the form
\begin{align}
z &= g_\phi(u, \epsilon) \\
u  &= f_\phi(x), \label{q:deepnet}
\end{align}
where $f_\phi$ and $g_\phi$ are surjective mappings, $\epsilon$ is noise drawn from
a prescribed probability distribution, $u$ is a vector, and $z$ is the feature vector.  In
practice we use the following simple and natural realization of the arrangement:
\begin{align}
z &=  f_\phi(x)  + \sigma_\epsilon \epsilon,
\label{q:reparameterization}
\end{align}
where $\epsilon \sim \mathcal{N}(0,I)$ is drawn from a zero-mean multivariate normal
distribution, with identity as covariance matrix, and $\sigma_\epsilon$ is a gain. The map
$f_\phi(\cdot)$ is implemented with a deep neural network, with its weights characterized
by $\phi$. For inference, $\sigma_\epsilon$ is set to zero. 

\subsection{Clone-based training}
\label{s:clones}

\begin{figure}[t]
\begin{center}
\includegraphics[width=0.4\textwidth]{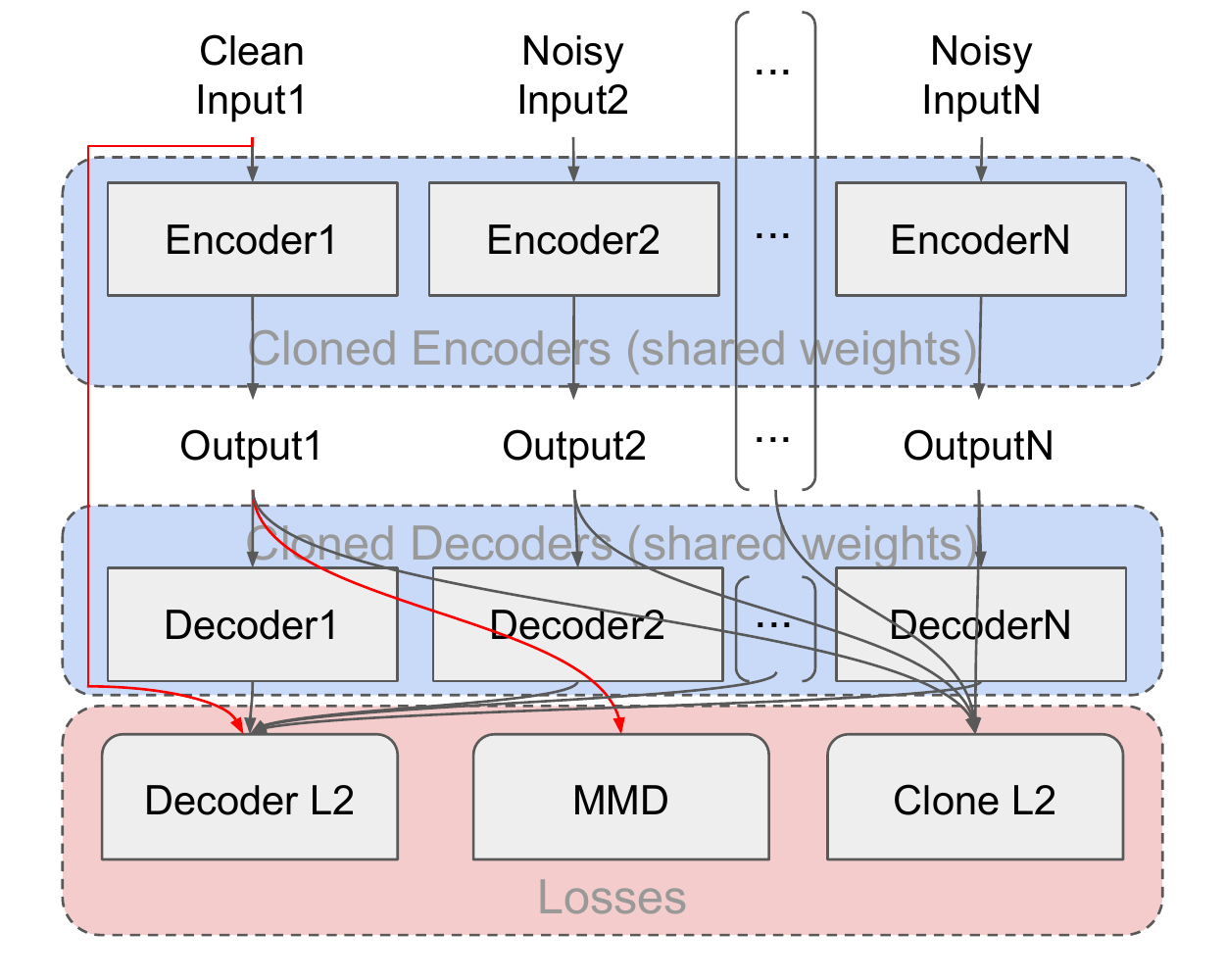}
\caption{Clone-based training setup. Red arrows hightlight when only the first
  clone is used. The decoders are optional.} 
\label{f:clonesetup}
\end{center}
\vspace{-2em}
\end{figure}

To allow the encoder network to learn about saliency, we employ clone-based
training, which is based on the Siamese network \cite{bromley1994signature} paradigm, but
may be endowed with a decoder. A diagram of the method is shown in Fig. \ref{f:clonesetup}.  Each
clone is an identical copy of a basic network. The clones are provided with different
input signals that are selected from a set of equivalent signals. The clone-based training
structure is used to find what information is shared between the different equivalent
signals and to represent that information in the form of features.  



To obtain an encoder with the  attributes given in section \ref{s:encoder}, 
we use an objective function with two or three components for clone-based training. A first
component encourages the similarity of the feature vectors $z^{(q)} \in \mathbb{R}^d$
across the clones $q \in \{1,\cdots,Q\}$. A second component  encourages that the
distribution of the feature vector is of a desired character. The optional third component encourages
the representation to provide high-fidelity decoding to a target signal, which typically
is derived from a clean signal.
 
Let $p_{Z^{(q)}}$ be the probability density of the random feature vector  $Z^{(q)}$ with
realization $z^{(q)}$ and let $p_{Z^{(1)},\cdots,   Z^{(Q)}}$ be the joint density of the
feature vectors of the clones. Let $\mathrm{E}[\cdot ]$ denote expectation over the input data
distribution, which can be approximated by suitable averaging over minibatches. The global
objective function is then  
  \begin{multline}
    D(p_{Z^{(1)},\cdots, Z^{(Q)}}) = \mathrm{E}[ D_s(Z^{(1)},\cdots, Z^{(Q)}) ]+ \\
   \sum_{i=1}^Q \left(  \lambda_f  D_{f}(p_{Z^{(i)}}) +
    \lambda_d \mathrm{E}[ D_{d}( h_\psi(Z^{(i)}), Y)] \right),
  \label{q:total}
    \end{multline}
where $D_s$ is a similarity measure for the clone feature sets,
$D_{f}$ is a measure on a distribution, $D_d$ is the optional measure of
decoder performance, $h_\psi$ is the decoder network with trained parameters $\psi$,  $Y$ is the desired output, and
the $\lambda_f$ and $\lambda_d$ are weightings. Note that the evaluation of $D_f$ and $D_d$ can be
limited to a smaller number of clones. Below, we discuss the individual terms of the objective
function in more detail.

A natural implementation of the first term of \eqref{q:total} is
\begin{align} 
D_s(z^{(1)},\cdots, z^{(Q)}) = \sum_{q=1}^Q \| z^{(1)} - z^{(q)} \|^2_2.
\label{q:norml}
\end{align}
An alternative is the 1-norm and we can add cross terms for all clones.
The shared features are found by encouraging the deterministic mapping $f_\phi$ to result
in outputs that are maximally similar for all clones, despite their different inputs. The
method preferably selects features describing information components
that are shared between the clone inputs at relatively high fidelity. For example, if the
clones receive noisy versions of segments of a speech signal, then a first feature may
describe a measure of spectral amplitude for a frequency region where the signal-to-noise
ratio is high. Note that this process naturally leads to \textit{disentanglement}.

Various approaches can be used to encourage the features to be independent, and
have a given variance, some requiring a desired distribution. Examples are the chi-square test, maximum mean discrepancy (MMD) 
\cite{gretton2007kernel,gretton2012kernel}, and the earth-moving distance reformulated
via the Kantorovich-Rubinstein duality, e.g.,
\cite{arjovsky2017wasserstein,frogner2015learning}.  Anticipating sparse features, we 
implement our system using MMD with an iid Laplacian desired distribution.

Let the desired iid Laplacian distribution be denoted as $p_{\mathcal{D}}$.  Let the
variable $\mathtt{V}$ be distributed as $\mathtt{V} \sim p_{\mathcal{D}}$.  We use as measure on the
distribution $p_{p_Z^{(1)}}$ the MMD between
the distributions $p_{Z^{(1)}}$ and $p_{\mathcal{D}}$:
\begin{multline} 
 D_{f}(p_{Z^{(1)}}) = \mathrm{MMD}(p_{Z^{(1)}}, p_{\mathcal{D}} )  \\
 =  \sup_{\omega \in \mathcal{H}, \,\,\|\omega\|_\mathcal{H} \leq 1} \mathrm{E}_{p_{Z^{(1)}}} (\omega(Z^{(1)})) - \mathrm{E}_{p_{\mathcal{D}}}(\omega(\mathtt{V})),
\label{q:mmd0}
\end{multline}
where $\mathcal{H}$ is a selected reproducing kernel Hilbert space (RKHS) and $\omega:
\mathbb{R}^d \rightarrow \mathbb{R}$ is a \textit{witness} function that is selected to
maximize the discrepancy subject to being in the unit ball in the RKHS. For a RKHS
associated with a kernel $k(\cdot,\cdot)$ an empirical representation of the square of
\eqref{q:mmd0} is \cite{gretton2007kernel,gretton2012kernel} 
\begin{multline} 
\mathrm{MMD}^2(\{z\},\{\mathtt{v}\})  = \frac{1}{M(M-1)} \\
 \sum_{i \neq j} k(z_i, z_j)  -  k(z_i, \mathtt{v}_j) - k(z_j, \mathtt{v}_i)  +  k(\mathtt{v}_i, \mathtt{v}_j), 
\label{q:mmd}
\end{multline}
where we omitted the superscript from $z^{(1)}$ to avoid notational clutter, where
$\{\cdot\}$ denotes a data batch, and where $M$ is the number of data in a batch.

The third term of \eqref{q:total} optimizes decoder clones $h_\psi(Z^{(i)})$ to match a target
output $Y$. Typically the target output is chosen to be a convenient representation of
the signal the features are desired to represent, with an appropriate criterion. For example, a mel-spectrum
representation of a clean speech signal with a squared error can be used.

To summarize the training procedure: we find the optimal deterministic function $f_\phi$
specifying the feature encoder by minimizing the objective function \eqref{q:total} with respect
to the encoder parameters $\phi$ and (optionally) the decoder parameters $\psi$.  The minimization is
performed using \eqref{q:norml}, \eqref{q:mmd}  and the distortion on the target outputs.

\section {Experimental Results}
\label{s:experiments}

This section provides results for a toy experiment and real-world data.
\vspace{-0.5em}
\subsection{System setup}

The same basic configuration is used for the toy and real-world experiments. We first
describe the real-world setup and then note the difference with the toy experiments.

For the real-world system the encoder network is
a stack of two LSTM layers \cite{hochreiter1997long} with residual connections, followed by
a fully connected layer (FC)
with residual connections. The LSTMs use ReLU activations and the FC layer used a tanh
for activation.  The input for each clone is a sequence of 80 or 240 mel bins that are
extracted from 40~ms windows the noisy signal, with a hop length of 20~ms.  A different
noise source is used for each of the clone encoders, mixed into 
the clean signal with 0-10~dB  signal-to-noise ratio (SNR). The clean utterance is kept
the same across clones.  The 
LSTM layers and penultimate FC layer had 800 units each.  The last layer is a fully
connected layer with a linear activation   
function. This last layer has 12 outputs, which corresponds to the components of $f_\phi(x)$ in
\eqref{q:reparameterization}.  The training used 32 clones and we used
$\lambda_f=1$ and $\lambda_d=18$.  The Adam optimizer \cite{kingma2014adam} was used
with a learning rate of 0.0001. 

Consistent with \eqref{q:reparameterization}, we add Gaussian noise to  $f_\phi(x)$ during the
training stage. The value of $\sigma_\epsilon$ is subject to a schedule that reduces its
value from 0.2 with an exponent of 0.98 per 1000 steps. For inference, we set
$\sigma_\epsilon =0$.

The toy experiment used the same configuration with the following differences. The encoder
consisted of three fully connected layers, an 80-dimensional input was used and the output
was two-dimensional.

Whenever a decoder was used, a decoder was added to each encoder. The decoder was
constructed to mirror the encoder. It consists of one fully connected layer, followed by
two LSTM layers, followed by a reshaping fully connected layer with linear activations to
match the input dimensionality. The output uses as criterion an 2-norm error measure.

\subsection{Toy experiment}
The toy experiment is aimed at showing that our feature extractor, \textit{without the optional
decoder,} can find and disentangle highly non-linear feature sets shared by the clone input signals. 

\vspace{-0.5em}
\subsubsection{Data generation}
The toy data approximate a spectral-domain description of whispered
speech. The objective of the feature extraction is to find the formants that form the
salient information shared across the clones. Let $P$ denote the number of formants
and let each formant contain $L$ signal components.  Let $V_{p,l}$ be a 
$d$-dimensional basis vector
representing a frequency (can be drawn from a multivariate Gaussian distribution). Let the uniformly
distributed scalar RV $\Psi_p(t) \sim  \mathcal{U}(0,1)$ at instance $t$ with realization
$\psi_p(t)$ be the formant gain and let the uniformly distributed scalar RV $W_{p,l}^{(q)}(t) \sim
\mathcal{U}(-1,1)$ with realization $w_{p,l}^{(q)}(t)$ be the corresponding excitation signal for
frequency $l$ of formant $p$ observed by clone $q$. Then we have
\vspace{-0.3em}
\begin{align}
x^{(q)}(t) = \sum_{p=1}^P \sum_{l=1}^L (\psi_p(t) + \gamma_p^{(q)}(t))\,  w^{(q)}_{p,l}(t)\, v_{p,l},
\end{align}
\vspace{-0.3em}
where the scalar RV $\Gamma_{p,l}^{(q)}(t) \sim \mathcal{U}(-b_{\Gamma_p}, b_{\Gamma_p})$ with
realization $\gamma_{p,l}^{(q)}(t)$ determines the variance of formant $p$. Importantly, the formants
$\psi_p(t)$ are identical for all clone inputs while the excitations $w_{p,l}^{(q)}(t)$ are different.

In the experiments, we set $P=2$, $d=30$, $L=10$, $b_{\Gamma_{1}}=0.01$ and
$b_{\Gamma_{2}}=0.005$. Note that the Laplacian desired distribution of \eqref{q:mmd0} is
not a good match. We used one million steps with a batch size of 144 for training.

\subsubsection{Results}

\begin{figure}[h] 
\vspace{-0.3cm}
\begin{center}
  \includegraphics[width=0.4\textwidth]{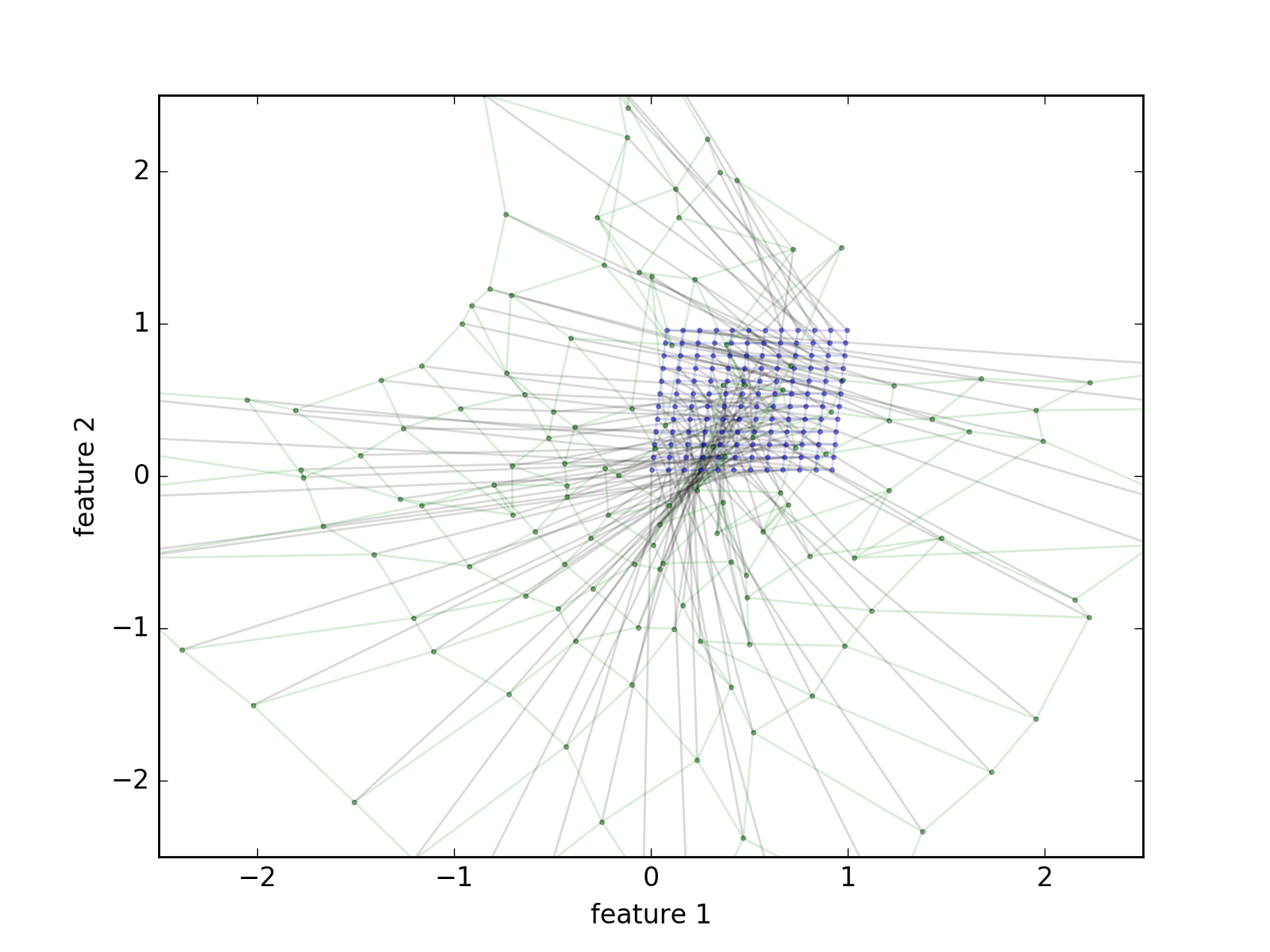}
  \hspace{0.01\textwidth}
\caption{Ground-truth features (blue) and extracted salient features (green) for the toy experiment.  Colored lines connect nodes that were neighbors in the ground truth, and black lines connect ground truth to extracted features.}
\label{f:toy1and2}
\end{center}
\vspace{-0.3cm}
\end{figure}

We provide a typical visual result for the toy experiment. Fig. \ref{f:toy1and2}
shows a mapping from ground-truth formants (blue) to the extracted salient features
(green). It is observed from the figure that the mapping is smooth and injective,  and
that the input features are distentangled. The clone-based successfully extracts the formant structure.
The results support that varying formant variance is important for disentanglement.
The measure \eqref{q:norml} encourages separation of formants with differing variance. 

We conclude from the toy experiments that \textit{i}) formants can be
extracted with the clone approach without a decoder and \textit{ii}) disentanglement
benefits from using the squared error criterion \eqref{q:norml} and from formants having
different variance (different $b_{\Gamma_{q}}$). This is true even if the groundtruth and
desired distribution used in MMD are mismatched.

\vspace{-1em}
\subsection{WaveNet experiments}
 
The WaveNet \cite{WaveNet:2016a} algorithm requires conditioning to produce natural-sounding
speech.  Our goal was to extract 12 salient features that can be used as conditioning to
produce clean speech signals from input that can be noisy or clean. 

\subsubsection{Databases and test procedure}

We used the 
WSJ0 database \cite{paul1992design} for training and
testing. The training set contained 32580 utterances by 123 speakers and the test set
contained 2907 utterances by 8 speakers. Additionally we used a mixed corpus of stationary and  
non-stationary noise from approximately 10,000 recordings captured in a variety of environments 
including busy streets, cafes, and pools. The inputs to the clones were 32 different
versions of a signal that contains an utterance, including the clean utterance and
versions with noise additions at an SNR of 0 to 10~dB.
When a decoder was used, the clean signal was used as target for all decoders.

To facilitate learning, the 16~kHz signals were pre-processed into an oversampled log mel spectrogram
representation. We considered two specific representations. The single window (sw)
approach uses 40~ms windows
with a time shift of 20 ms and 80 log mel coefficients for each time shift. In the dual
window (dw) approach each 20~ms shift is associated with one window of 40 ms and two
windows of 20~ms  (located at 5-25, and 15-45~ms of the 40~ms window). The 20~ms windows
were described with  80 log mel spectrogram coefficients, for a total of 240 coefficients
for each 20~ms shift. 

As reference system we used feature sets obtained with principal component analysis (PCA).
It extracted 12 features (PCA12) from the 240-dimensional vector of the dual window data. The
PCA was computed for the input signals of the clone-based training.
Additionally, a PCA that extracts four features (PCA4) was trained as an anchor for the listening test.

We conducted a MUSHRA-like listening test on 10 utterances from the test set for 10
different speakers, using 100 human raters per utterance.  The MUSHRA reference was the
clean utterance. We evaluated the output of each model for both clean and noisy test
utterances, except for PCA4.

\subsubsection{Results}

\begin{figure}[t]
\begin{center}
\includegraphics[width=0.4\textwidth]{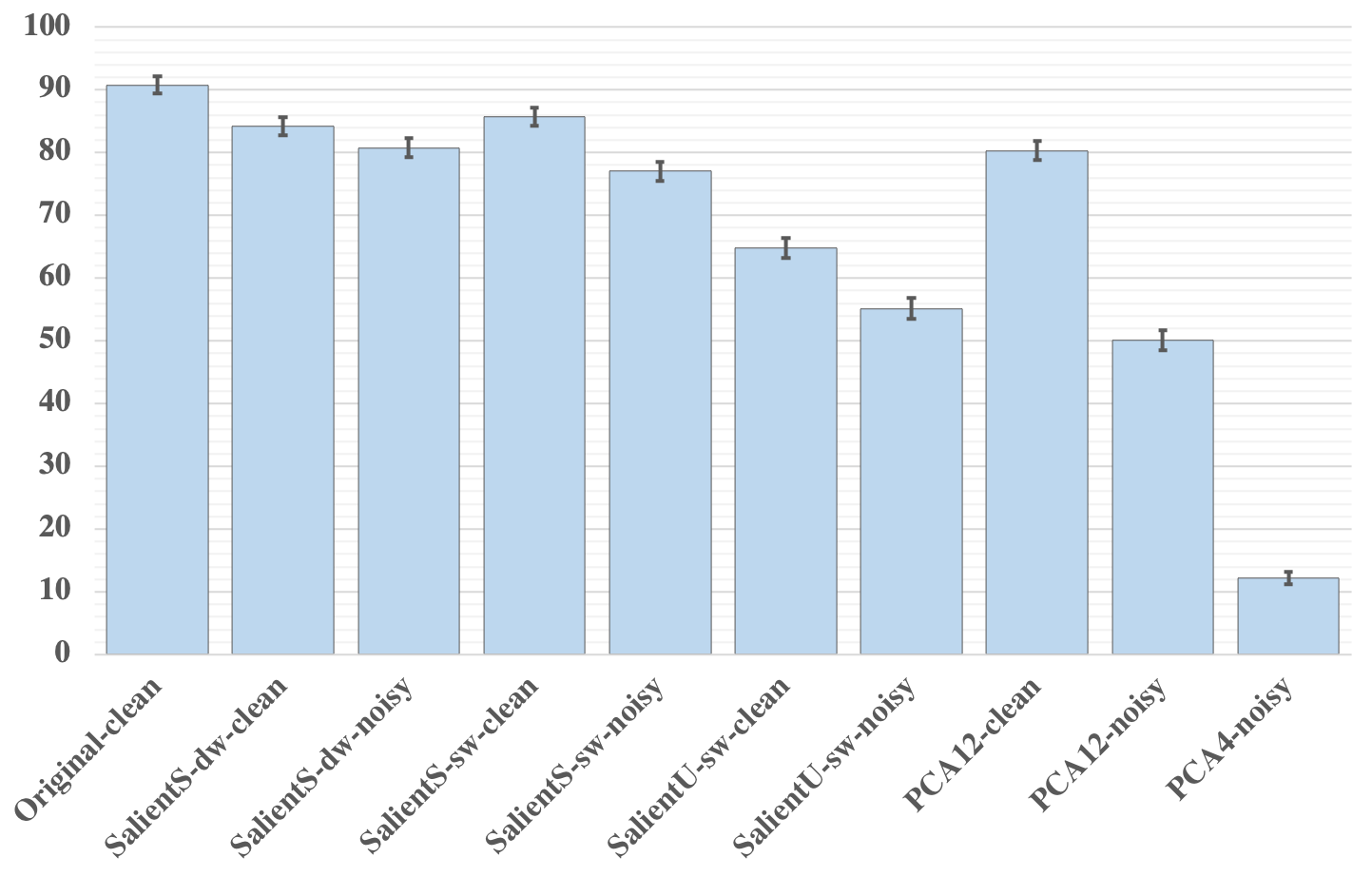}
\caption{Listening test results.} 
\label{f:listening}
\end{center}
\vspace{-1.8em}
\end{figure}

Fig. \ref{f:listening} shows the result of the listening test. 
Clone-based learning without a decoder with a single window (SalientU-sw)
provides natural speech quality with good speaker identity but with fairly frequent errors
for phonemes of short duration. As expected, the number of errors is lower for the clean
(SalientU-sw-clean) than for noisy input signals. Learning with a decoder with a single window
(SalientS-sw) reduces the errors to being very infrequent with a further small improvement
for the clean input signals (SalientS-sw-clean). For the noisy input the errors are further
reduced by using the dual window approach (SalientS-dw-noisy), reaching almost the quality
obtained with clean input (SalientS-dw-clean and SalientS-sw-clean).

In summary, the clone-based systems with decoder significantly outperformed the reference
system. This is particularly true under noisy conditions (SalientS-dw vs PCA12).

\section {Conclusion}
\label{s:conclusion}

We showed that clone-based training allows saliency to be defined in a qualitative manner
by a system designer. From experiments with a toy example, we conclude that clone-based 
training can be used to disentangle formants from a signal. In a real-world application,
the addition of a decoder improved the performance of the clone-based feature-extraction
system further. The clone-based system is inherently robust to distortion and it
significantly outperformed a reference system. Its natural application is coding and
enhancement. 
\clearpage

\bibliographystyle{IEEEtran}
\bibliography{Refs_Enhancement,Refs_BeamFormer,Refs_Coding,Refs_Autoencoder}

\begin{thebibliography}{10}
\providecommand{\url}[1]{#1}
\csname url@samestyle\endcsname
\providecommand{\newblock}{\relax}
\providecommand{\bibinfo}[2]{#2}
\providecommand{\BIBentrySTDinterwordspacing}{\spaceskip=0pt\relax}
\providecommand{\BIBentryALTinterwordstretchfactor}{4}
\providecommand{\BIBentryALTinterwordspacing}{\spaceskip=\fontdimen2\font plus
\BIBentryALTinterwordstretchfactor\fontdimen3\font minus
  \fontdimen4\font\relax}
\providecommand{\BIBforeignlanguage}[2]{{%
\expandafter\ifx\csname l@#1\endcsname\relax
\typeout{** WARNING: IEEEtran.bst: No hyphenation pattern has been}%
\typeout{** loaded for the language `#1'. Using the pattern for}%
\typeout{** the default language instead.}%
\else
\language=\csname l@#1\endcsname
\fi
#2}}
\providecommand{\BIBdecl}{\relax}
\BIBdecl

\bibitem{chen2016infogan}
X.~Chen, Y.~Duan, R.~Houthooft, J.~Schulman, I.~Sutskever, and P.~Abbeel,
  ``Infogan: Interpretable representation learning by information maximizing
  generative adversarial nets,'' in \emph{Advances in Neural Information
  Processing Systems}, 2016, pp. 2172--2180.

\bibitem{kingma2013auto}
D.~P. Kingma and M.~Welling, ``Auto-encoding variational {Bayes},'' \emph{arXiv
  preprint arXiv:1312.6114}, 2013.

\bibitem{rezende2014stochastic}
D.~J. Rezende, S.~Mohamed, and D.~Wierstra, ``Stochastic backpropagation and
  approximate inference in deep generative models,'' \emph{arXiv preprint
  arXiv:1401.4082}, 2014.

\bibitem{chen2016variational}
X.~Chen, D.~P. Kingma, T.~Salimans, Y.~Duan, P.~Dhariwal, J.~Schulman,
  I.~Sutskever, and P.~Abbeel, ``Variational lossy autoencoder,'' \emph{arXiv
  preprint arXiv:1611.02731}, 2016.

\bibitem{zhao2017infovae}
S.~Zhao, J.~Song, and S.~Ermon, ``{InfoVAE}: Information maximizing variational
  autoencoders,'' \emph{arXiv preprint arXiv:1706.02262}, 2017.

\bibitem{alemi2017information}
A.~A. Alemi, B.~Poole, I.~Fischer, J.~V. Dillon, R.~A. Saurous, and K.~Murphy,
  ``An information-theoretic analysis of deep latent-variable models,''
  \emph{arXiv preprint arXiv:1711.00464}, 2017.

\bibitem{PhuongMutualA}
M.~Phuong, M.~Welling, N.~Kushman, R.~Tomioka, and S.~Nowozin, ``The mutual
  autoencoder: Controlling information in latent code representations,'' 2017,
  https://openreview.net/forum?id=HkbmWqxCZ.

\bibitem{Huszar2017Ismax}
F.~Huszar, ``Is maximum likelihood useful for representation learning?'' 2017,
  ttp://www.inference.vc/maximum-likelihood-for-representation-learning-2/.

\bibitem{braithwaite2018bounded}
D.~Braithwaite and W.~B. Kleijn, ``Bounded information rate variational
  autoencoders,'' \emph{arXiv preprint arXiv:1807.07306}, 2018.

\bibitem{bengio2013representation}
Y.~Bengio, A.~Courville, and P.~Vincent, ``Representation learning: A review
  and new perspectives,'' \emph{IEEE transactions on pattern analysis and
  machine intelligence}, vol.~35, no.~8, pp. 1798--1828, 2013.

\bibitem{louizos2015variational}
C.~Louizos, K.~Swersky, Y.~Li, M.~Welling, and R.~Zemel, ``The variational fair
  autoencoder,'' \emph{arXiv preprint arXiv:1511.00830}, 2015.

\bibitem{higgins2017beta}
I.~Higgins, L.~Matthey, A.~Pal, C.~Burgess, X.~Glorot, M.~Botvinick,
  S.~Mohamed, and A.~Lerchner, ``$\beta$-{VAE}: Learning basic visual concepts
  with a constrained variational framework,'' in \emph{International Conference
  on Learning Representations}, 2017.

\bibitem{burgess2018understanding}
C.~P. Burgess, I.~Higgins, A.~Pal, L.~Matthey, N.~Watters, G.~Desjardins, and
  A.~Lerchner, ``Understanding disentangling in $\beta$-{VAE},'' \emph{arXiv
  preprint arXiv:1804.03599}, 2018.

\bibitem{esmaeili2018structured}
B.~Esmaeili, H.~Wu, S.~Jain, A.~Bozkurt, N.~Siddharth, B.~Paige, D.~H. Brooks,
  J.~Dy, and J.-W. van~de Meent, ``Structured disentangled representations,''
  \emph{stat}, vol. 1050, p.~29, 2018.

\bibitem{kim2018disentangling}
H.~Kim and A.~Mnih, ``Disentangling by factorising,'' \emph{arXiv preprint
  arXiv:1802.05983}, 2018.

\bibitem{tishby2000information}
N.~Tishby, F.~C. Pereira, and W.~Bialek, ``The information bottleneck method,''
  \emph{arXiv preprint physics/0004057}, 2000.

\bibitem{alemi2016deep}
A.~A. Alemi, I.~Fischer, J.~V. Dillon, and K.~Murphy, ``Deep variational
  information bottleneck,'' \emph{arXiv preprint arXiv:1612.00410}, 2016.

\bibitem{kleijn2017wavenet}
W.~B. Kleijn, F.~S. Lim, A.~Luebs, J.~Skoglund, F.~Stimberg, Q.~Wang, and T.~C.
  Walters, ``{WaveNet} based low rate speech coding,'' \emph{arXiv preprint
  arXiv:1712.01120}, 2017.

\bibitem{hecht2005extraction}
R.~M. Hecht and N.~Tishby, ``Extraction of relevant speech features using the
  information bottleneck method,'' in \emph{Ninth European Conference on Speech
  Communication and Technology}, 2005.

\bibitem{bromley1994signature}
J.~Bromley, I.~Guyon, Y.~LeCun, E.~S{\"a}ckinger, and R.~Shah, ``Signature
  verification using a "{Siamese}" time delay neural network,'' in
  \emph{Advances in Neural Information Processing Systems}, 1994, pp. 737--744.

\bibitem{hoffer2015deep}
E.~Hoffer and N.~Ailon, ``Deep metric learning using triplet network,'' in
  \emph{International Workshop on Similarity-Based Pattern Recognition}.\hskip
  1em plus 0.5em minus 0.4em\relax Springer, 2015, pp. 84--92.

\bibitem{kamper2016deep}
H.~Kamper, W.~Wang, and K.~Livescu, ``Deep convolutional acoustic word
  embeddings using word-pair side information,'' in \emph{Acoustics, Speech and
  Signal Processing (ICASSP), 2016 IEEE International Conference on}.\hskip 1em
  plus 0.5em minus 0.4em\relax IEEE, 2016, pp. 4950--4954.

\bibitem{WaveNet:2016a}
A.~{van den Oord}, S.~{Dieleman}, H.~{Zen}, K.~{Simonyan}, O.~{Vinyals},
  A.~{Graves}, N.~{Kalchbrenner}, A.~{Senior}, and K.~{Kavukcuoglu},
  ``{WaveNet: A Generative Model for Raw Audio},'' \emph{ArXiv e-prints}, Sep.
  2016.

\bibitem{gretton2007kernel}
A.~Gretton, K.~M. Borgwardt, M.~Rasch, B.~Sch{\"o}lkopf, and A.~J. Smola, ``A
  kernel method for the two-sample-problem,'' in \emph{Advances in neural
  information processing systems}, 2007, pp. 513--520.

\bibitem{gretton2012kernel}
A.~Gretton, K.~M. Borgwardt, M.~J. Rasch, B.~Sch{\"o}lkopf, and A.~Smola, ``A
  kernel two-sample test,'' \emph{Journal of Machine Learning Research},
  vol.~13, no. Mar, pp. 723--773, 2012.

\bibitem{arjovsky2017wasserstein}
M.~Arjovsky, S.~Chintala, and L.~Bottou, ``Wasserstein generative adversarial
  networks,'' in \emph{International Conference on Machine Learning}, 2017, pp.
  214--223.

\bibitem{frogner2015learning}
C.~Frogner, C.~Zhang, H.~Mobahi, M.~Araya, and T.~A. Poggio, ``Learning with a
  {Wasserstein} loss,'' in \emph{Advances in Neural Information Processing
  Systems}, 2015, pp. 2053--2061.

\bibitem{hochreiter1997long}
S.~Hochreiter and J.~Schmidhuber, ``Long short-term memory,'' \emph{Neural
  computation}, vol.~9, no.~8, pp. 1735--1780, 1997.

\bibitem{kingma2014adam}
D.~P. Kingma and J.~Ba, ``Adam: A method for stochastic optimization,''
  \emph{arXiv preprint arXiv:1412.6980}, 2014.

\bibitem{paul1992design}
D.~B. Paul and J.~M. Baker, ``The design for the wall street journal-based csr
  corpus,'' in \emph{Proceedings of the workshop on Speech and Natural
  Language}.\hskip 1em plus 0.5em minus 0.4em\relax Association for
  Computational Linguistics, 1992, pp. 357--362.

\end{thebibliography}

\end{document}